\documentclass[usenatbib]{mnras}

\usepackage{graphicx}
\usepackage{amsmath,amsfonts,amssymb}
\usepackage{multicol}

\newcommand{\be}{\begin{equation}} 
\newcommand{\ee}{\end{equation}}
\newcommand{\bea}{\begin{equation}\begin{aligned}} 
\newcommand{\eea}{\end{aligned}\end{equation}}

\newcommand{\td}{{\rm d}}
\newcommand{\vect}[1]{\boldsymbol{#1}}

\newcommand{\nobreakfootnote}[1]{%
  \footnote{\begin{minipage}[t]{\linewidth}#1\end{minipage}}%
}

\pubyear{2026}
\volume{547}

\makeatletter
\let\old@maketitle\@maketitle

\def\@maketitle{%
  \begingroup
  \let\@journal\@empty
  \let\@volume\@empty
  \let\@pagerange\@empty
  \old@maketitle
  \endgroup
}
\makeatother

\title[Weak lensing of bright standard sirens]{Weak lensing of bright standard sirens: prospects for $\sigma_8$}

\author[V. Vaskonen]{
Ville Vaskonen\thanks{ville.vaskonen@pd.infn.it}
\\
Dipartimento di Fisica e Astronomia, Universit\`a degli Studi di Padova, Via Marzolo 8, 35131 Padova, Italy \\
Istituto Nazionale di Fisica Nucleare, Sezione di Padova, Via Marzolo 8, 35131 Padova, Italy \\
Keemilise ja bioloogilise f\"u\"usika instituut, R\"avala pst. 10, 10143 Tallinn, Estonia
}

\begin{document}

\label{firstpage}
\pagerange{\pageref{firstpage}--\pageref{lastpage}}
\maketitle

\begin{abstract}
Gravitational wave events with electromagnetic counterparts provide direct measurements of the Hubble diagram. We demonstrate that incorporating weak lensing into bright standard siren analyses allows measurements of cosmological parameters that do not influence the mean luminosity distance-redshift relation but do impact the cosmic structures. In particular, we examine the prospects for measuring the standard deviation of matter perturbations $\sigma_8$, in addition to the Hubble constant and the matter abundance. We find that a $10\%$ measurement of $\sigma_8$ would be feasible with ET, provided a population of $300$ neutron star binaries with electromagnetic counterparts is observed. With LISA, the measurement of $\sigma_8$ would have $30\%$ accuracy, assuming a population of $12$ massive black hole binaries with electromagnetic counterparts is observed.
\end{abstract}


\section{Introduction}

Gravitational waves (GWs) from compact binary coalescences provide a unique avenue to probe cosmic expansion through the construction of a Hubble diagram that is independent of the traditional cosmic distance ladder. GW events with electromagnetic (EM) counterparts act as bright standard sirens, allowing direct measurements of luminosity distances from the waveform amplitude. While standard candles have established the accelerated expansion of the Universe~\citep{SupernovaSearchTeam:1998fmf,SupernovaCosmologyProject:1998vns}, bright sirens offer independent constraints on the Hubble constant and other cosmological parameters~\citep{Schutz:1986gp,Holz:2005df,LIGOScientific:2017adf,Jin:2025dvf}.

The propagation of GWs across cosmological distances is affected by intervening cosmic structures through gravitational lensing. Searches for such signatures are actively underway within the LIGO-Virgo-KAGRA (LVK) network. Analyses from the fourth observing run have reported no definitive detections, although an intriguing candidate, GW231123, has been identified~\citep{Chan:2025kyu,Goyal:2025eqo,Chakraborty:2025pxt,Shan:2025dcd}. These early studies highlight the increasing importance of accounting for lensing systematics as detector sensitivities improve and event catalogues grow.

In the weak-lensing regime, gravitational lensing results in the magnification or demagnification of the observed strain by a factor of $\sqrt{\mu}$ relative to the unlensed signal. The resulting scatter in luminosity distances is characterized by the magnification probability distribution $\td P/\td \mu$, which is non-Gaussian and evolves with redshift. At the redshifts probed by third-generation ground-based detectors such as ET~\citep{ET:2025xjr} and space-based detectors such as LISA~\citep{LISA:2017pwj}, weak-lensing scatter will contribute percent-level uncertainties to individual distance measurements and must therefore be incorporated carefully into cosmological analyses.

Accurate modeling of $\td P/\td \mu$ is essential for robust cosmological inference with bright sirens, as simplified treatments can bias parameter estimates and underestimate true uncertainties~\citep{Fleury:2016fda,Adamek:2018rru,Cusin:2020ezb,Canevarolo:2023dkh,Mpetha:2024xiu}. In addition, $\td P/\td \mu$ encodes information about matter inhomogeneities. Measurements of the scatter in the Hubble diagram can thus provide insights into cosmology and the properties of cosmic structures. In particular, the scatter is sensitive to the standard deviation of matter perturbations, $\sigma_8$. The potential of lensing-induced scatter to constrain $\sigma_8$ has been explored for standard candles by~\citet{Quartin:2013moa,Castro:2014oja}. Recent results from the Dark Energy Survey demonstrate a $>5\sigma$ detection of weak-lensing magnification of Type Ia supernovae~\citep{DES:2024lto}, confirming that Hubble diagram scatter can indeed serve as a probe of $\sigma_8$.

The possibility of constraining $\sigma_8$ through weak lensing of bright sirens was first explored by~\citet{Congedo:2018wfn}, where the lensing signal was characterized via the convergence power spectrum computed at the linear level. This approach was subsequently extended to a larger set of cosmological parameters and dark sirens by~\citet{Mpetha:2022xqo}, and to cross-correlations with galaxy fields by~\citet{Balaudo:2022znx}. The linear convergence power spectrum is valid only on large scales, where a few hundred sources are required for a reliable measurement. With a sample of $\mathcal{O}(10^4)$ sources, a $3\%$ measurement of $\sigma_8$ was predicted by~\citet{Congedo:2018wfn}.

In this work, we instead exploit the full magnification distribution, incorporating modeling of non-linear structures. This approach, first mentioned by~\citet{Congedo:2018wfn}, is complementary to that based on the convergence power spectrum and captures additional information, including lensing of individual sources and non-Gaussianity, allowing better sensitivity to $\sigma_8$. Indeed, we find that ET could measure $\sigma_8$ to $10\%$ with $300$ sources, and LISA could reach $30\%$ with just $12$ sources. Modeling non-linear structures, however, introduces theoretical uncertainties~\citep{Mpetha:2024xiu,Alfradique:2024fkb}, and further work beyond the scope of this study is needed to refine the magnification distribution. The goal of this work is to demonstrate the prospects of this approach by implementing, for the first time, the magnification distribution in Hubble-diagram reconstruction. In addition to $\sigma_8$, we consider the Hubble parameter $h$ and the matter abundance $\Omega_M$, that primarily affect the mean luminosity distance-redshift relation.

\section{Lensing magnification}

We compute the lensing magnification distribution $\td P/\td \mu$ using the stochastic approach developed by~\citet{Kainulainen:2009dw,Kainulainen:2010at,Kainulainen:2011zx}, which we have implemented in \texttt{C++}.\nobreakfootnote{The full code to generate the magnification distribution and reconstruct the Hubble diagram is publicly available at~\href{https://github.com/vianvask/halos}{https://github.com/vianvask/halos}.} In this approach, the matter distribution along the line of sight is modeled as an ensemble of discrete lenses drawn from a specified mass function. For each realization, lenses are placed in redshift slices near the line of sight to the source, and the total magnification is obtained by summing the individual lensing contributions under the weak-lensing approximation. Repeating this procedure over many realizations yields the magnification probability distribution. This method is computationally efficient and significantly faster than approaches based on ray tracing through universes generated, for example, with N-body simulations~\citep{Holz:1997ic,Holz:2004xx,Takahashi:2011qd,Takahashi:2017hjr}. Furthermore, it has been argued e.g. by~\citet{DES:2025omh} that N-body simulations do not have sufficient resolution to accurately reconstruct the lensing magnification distribution for point sources.

The lensing magnification, defined as the ratio of the lensed to unlensed flux, is given by the convergence $\kappa$ and the shear $\gamma = \sqrt{\gamma_1^2+\gamma_2^2}$ as
\be \label{eq:mueq}
	\mu = \frac{1}{(1-\kappa)^2 - \gamma^2} \,.
\ee
In the weak-lensing approximation, $\kappa$, $\gamma_1$ and $\gamma_2$ of multiple lenses are obtained by summing the contributions of individual lenses:\nobreakfootnote{We retain the non-linear relation~\eqref{eq:mueq} to capture high-magnification events, which are dominated by single lenses. Our approach smoothly connects this regime to low magnifications, where linear superposition of $\kappa$ and $\gamma_{1,2}$ is adequate.} $\kappa = \sum_j \kappa_\epsilon(\vect\theta_j)$, $\gamma_1 = \sum_j \gamma_\epsilon(\vect\theta_j) \cos\phi_j$ and $\gamma_2 = \sum_j \gamma_\epsilon(\vect\theta_j) \sin\phi_j$, where $\phi_j$ denotes the polar angle of the lens in the lens plane, defined with respect to a coordinate system common to all lenses, and $\vect\theta_j$ encodes the lens redshift $z_j$, its projected distance $r_j$ from line of sight to the source, and parameters describing the lens properties and orientation. We include lensing contributions from both halos and filaments.

We model halos using the pseudo-elliptical Navarro-Frenk-White (NFW) profile, for which the expressions for $\kappa$ and $\gamma$ are derived by~\citet{Golse:2001ar}. The ellipticity is defined as $\epsilon = 1 - b/a$, where $a$ and $b$ are the semi-major and semi-minor axes of the lensing potential projected onto the lens plane. The ellipticity $\epsilon$, the scale radius $r_s$, and the characteristic density $\rho_s$ at the scale radius are functions of the halo mass $M$. The halo mass is defined as $M = (4\pi/3)\,\Delta\,\rho_c\,r_\Delta^3$, where $r_\Delta = C\,r_s$ and $\rho_c$ denotes the critical density. We adopt $\Delta = 200$, use the concentration-mass relation $C = C(M,z)$ from~\citet{Dutton:2014xda}, and the ellipticity fit $\epsilon = \epsilon(M,z)$ from~\citet{Allgood:2005eu}. Consequently, halos are characterized by the parameter set $\boldsymbol{\theta}_j = \{z_j, r_j, M_j, \phi_{e,j}\}$, where $\phi_{e,j}$ denotes the orientation of the halo ellipticity.

We model filaments as cylindrical structures with a radial density profile $\rho \propto (1 + r^2/r_s^2)^{-1}$, where $r_s$ denotes the filament scale radius. The filament mass is defined as $M = \pi r_s^2 L\,\bar{\rho}(r<r_s)/2$, where $L$ is the filament length and $\bar{\rho}(r<r_s)$ is the average density within radius $r_s$. Motivated by the results of~\citet{Colberg:2004cd}, we fix $r_s = 1\,\mathrm{Mpc}\,(M/10^{14}\,M_\odot)^{1/3}$, $L = 20\,\mathrm{Mpc}\,(M/10^{14}\,M_\odot)^{1/3}$, and $\bar{\rho}(r<r_s) = 10\,\rho_c$. Filaments are characterized by the set of parameters $\boldsymbol{\theta}_j = \{z_j, r_j, M_j, \theta_{f,j}, \phi_{f,j}\}$, where the angles $\theta_{f,j}$ and $\phi_{f,j}$ determine the orientation of the filament with respect to the lens plane.

\begin{figure}
\centering
\includegraphics[width=\columnwidth]{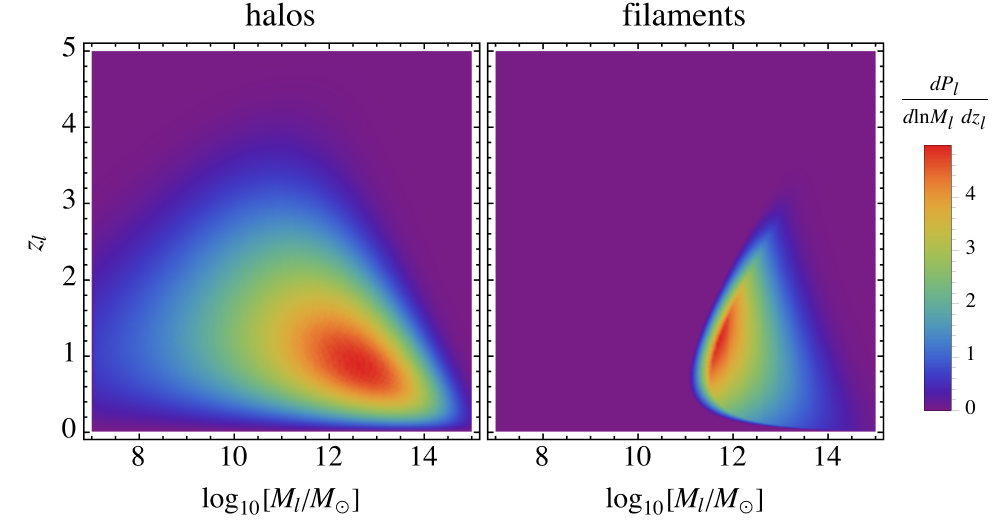}
\vspace{-4mm}
\caption{The PDF of lenses for sources at $z_s = 5$, integrated over the transverse distance $r$. In the purple region $r_{\rm max} = 0$.}
\label{fig:Plzs5}
\end{figure}

We estimate the halo and filament mass functions using the excursion-set formalism~\citep{Bond:1990iw}, adopting the mass-dependent collapse barriers presented by~\citet{Shen:2005wd}. The average comoving number density of potential lenses is given by
\be
    \frac{\td \bar{n}}{\td \ln M} = \frac{\rho_m}{M} A \left[ 1+(q\nu)^{-p} \right] \sqrt{\frac{q\nu}{2\pi}} e^{-q\nu/2} \frac{\td \ln \nu}{\td \ln M} \,,
\ee
where $\rho_m$ denotes the average present-day mass density and $A = [1 + 2^{-p} \Gamma(1/2 - p)/\sqrt{\pi}]^{-1}$ ensures normalization. We adopt $\{p,q\} = \{0.3, 0.8\}$ for halos and $\{p,q\} = \{0, 0.7\}$ for filaments.\nobreakfootnote{The values of $p$ and $q$ were determined by performing random walks and fitting the first-crossing distribution, following~\citet{Sheth:2001dp}.} The parameter $\nu$ is defined as $\nu \equiv \delta_c(z)^2/\sigma_M^2$, where $\delta_c(z)$ is the density contrast threshold for spherical collapse and $\sigma_M^2$ is the variance of the matter fluctuations at the mass scale $M$. We compute the latter using the real space top-hat window function and the matter power spectrum estimated as~\citet{Eisenstein:1997ik}. We fix the normalization by setting the value of $\sigma_8 = \sigma_M(R = 8h \,{\rm Mpc})$. 

We also include the linear bias factor $B(M_b,z) = 1 + (q \nu-1)/\delta_c(0) + 2p/[\delta_c(0)(1+(q\nu)^p)]$ with  $\{p,q\} = \{0.3, 0.75\}$ (see, e.g.~\citet{Baumann:2022mni}). The halo number density is modeled as a biased lognormal field $\td n = \td \bar{n} \exp[B(M_b,z)\delta_b - B(M_b,z)^2 \sigma_{M_b}^2/2]$ where $\delta_b$ is a Gaussian random variable with variance $\sigma_{M_b}^2$ and the mass scale $M_b$ corresponds to the average mass enclosed within a radius $r_{\rm max}$ defined below and within a redshift bin $\Delta z$, chosen so that $M_b$ is much larger than the typical lens masses.

The probability density function (PDF) of the lenses for a source at redshift $z_s$ is given by
\be \label{eq:Pl}
    \frac{\td P_l}{\td\ln M \td z \td r} \propto \frac{2\pi (1+z)^2 r}{H(z)} \frac{\td n}{\td \ln M} \theta(r_{\rm max}-r) \theta(z_s-z) ,
\ee
where $H(z)$ denotes the Hubble rate and $\theta$ the Heaviside function. We consider a flat $\Lambda$CDM model where
\be
    H(z) = H_0  \sqrt{\Omega_R (1+z)^4 + \Omega_M (1+z)^3 + \Omega_\Lambda} \,,
\ee
with $H_0 = 100h \,{\rm km}/{\rm s}/{\rm Mpc}$. We impose an upper limit $r_{\rm max} = r_{\rm max}(M,z)$ on the transverse distance $r$ of the lens from the line of sight to the source by requiring that for each individual lens $\kappa$ exceeds a threshold value $\kappa_{\rm thr}$ that we choose so that the average number of lens halos is $\bar{N}_{l,h}(r<r_{\rm max}) = 100$.\nobreakfootnote{This choice is made to balance computational efficiency and accuracy.} In Fig.~\ref{fig:Plzs5} we show the PDF of the lenses for the sources at $z_s = 5$, integrated over $r$. The threshold in this case is $\kappa_{\rm thr} \approx 0.0007$ and lenses in the purple region would give $\kappa < \kappa_{\rm thr}$. 

We have not taken into account the wave optics effects that suppress the magnification for~\citep{Fairbairn:2022xln,Urrutia:2024pos}
\be
	f \lesssim 3\times 10^{-6} \,{\rm Hz}\, \frac{D_{L,A} D_{LS,A}}{(1+z_l) D_{S,A} {\rm Gpc}} \left[ \frac{\rm kpc}{r} \right]^2 \,.
\ee
Integrating over the PDF of lenses, Eq.~\eqref{eq:Pl}, with $r_{\rm max}$ determined by the above condition, we find that the probability of having a halo with a mass exceeding $10^7\,M_\odot$ in the wave optics regime is less than $10^{-5} [f/{\rm Hz}]^{-1}$. Since halos lighter than $10^7\,M_\odot$ do not contribute significantly to weak lensing, as seen from Fig.~\ref{fig:Plzs5}, and the frequencies we consider are $f \gtrsim 0.1\,{\rm mHz}$, neglecting the wave optics effects appears reasonable.

\begin{figure}
\centering
\includegraphics[width=\columnwidth]{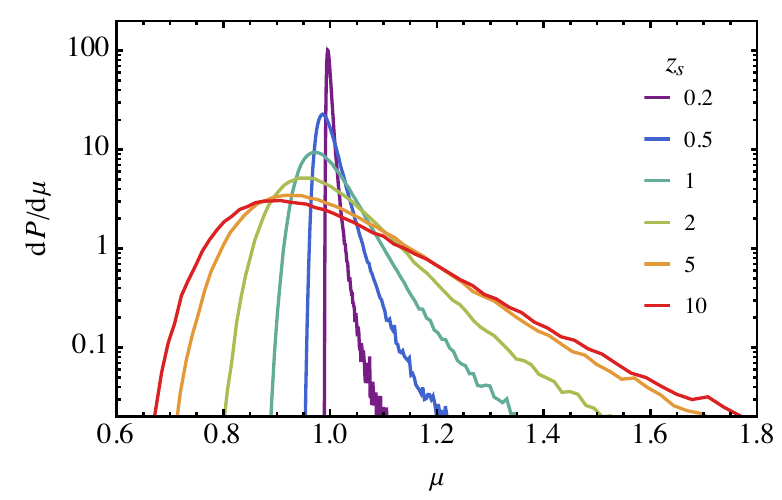}
\vspace{-4mm}
\caption{The PDF of lensing magnifications for different source redshifts for the Planck 2018 values of the cosmological parameters.}
\label{fig:dPdmu}
\end{figure}

We compute the distribution of lensing magnifications, $\mu$, by generating realizations of halos and filaments along the line of sight to a source at redshift $z_s$. For each realization $i$ and redshift bin, we draw the number of lenses from a Poisson distribution whose mean we compute accounting for the linear bias. We then sample the lens parameters from the probability distribution function in Eq.~\eqref{eq:Pl} and calculate the corresponding convergence $\kappa_i$ and shear $\gamma_i$. Next, we subtract the ensemble mean of $\{\kappa_i\}$ from each realization and compute the magnification $\mu_i$. We bin the magnifications and apply a $1/\mu$ factor to convert the image-plane distribution into the source-plane distribution. The resulting probability distribution functions of the lensing magnification for different source redshifts $z_s$, obtained from $4\times10^5$ realizations at each $z_s$, are shown in Fig.~\ref{fig:dPdmu} for the cosmological parameters inferred from CMB observations ($h = 0.674$, $\Omega_M = 0.315$, $\sigma_8 = 0.811$)~\citep{Planck:2018vyg}, which we adopt as our benchmark cosmology.

In a homogeneous and isotropic universe, the luminosity distance-redshift relation is given by
\be
    \tilde{D}_L(z) = (1+z) \int_0^z \frac{\td z’}{H(z’)} \,.
\ee
Gravitational lensing by intervening structure induces fluctuations around this background relation. The lensing magnification $\mu$ modifies the apparent luminosity distances according to
\be
    D_L(z_s) = \frac{\tilde{D}_L(z_s)}{\sqrt{\mu}} \,.
\ee

\begin{figure}
\centering
\includegraphics[width=\columnwidth]{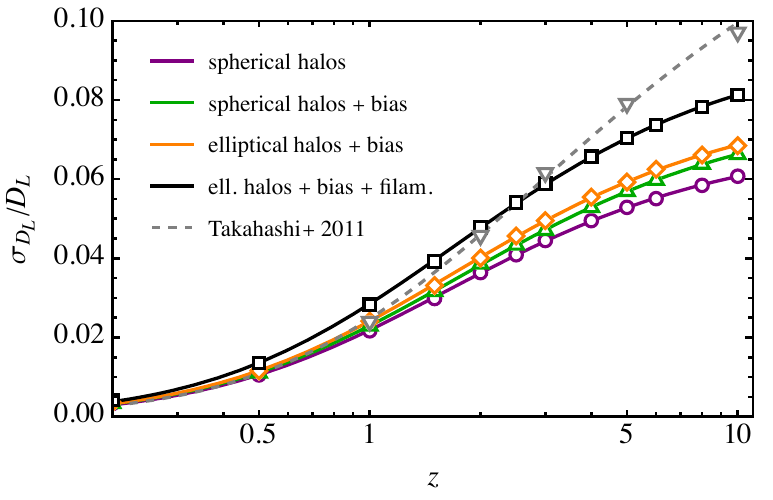}
\vspace{-4mm}
\caption{The variance of luminosity distances induced by weak lensing in various approximations of the lens population. For comparison, the dashed curve shows the result of~\citet{Takahashi:2011qd}. The open symbols depict the numerical results, while the curves illustrate the fit using the ansatz~\eqref{eq:sigmaDfit}.}
\label{fig:sigmaD}
\end{figure}

In Fig.~\ref{fig:sigmaD}, we show the variance of apparent luminosity distances as a function of redshift.\nobreakfootnote{We note that the weak lensing variance is often given in terms of apparent magnitude $m = 5 \log_{10} D_L + {\rm const.}$ (see e.g.~\cite{Marra:2013roi}). The relation between weak lensing scatter in $m$ and in $D_L$ is given by $\sigma_{D_L}/D_L 
\approx 0.46 \sigma_m$.} The open symbols show the results obtained by sampling $\mu$ from the magnification PDFs presented in Fig.~\ref{fig:dPdmu}. The curves show the fitting function~\citep{Hirata:2010ba}
\be \label{eq:sigmaDfit}
	\frac{\sigma_{D_L}}{D_L} = \frac{c}{2} \left[ \frac{1 - (1 + z)^{-\beta}}{\beta} \right]^\alpha \,.
\ee
Our full model, including elliptical halos, bias and filaments, gives $c = 0.22$, $\beta = 1.09$, $\alpha = 1.87$. We find reasonably good agreement with the results of~\citet{Takahashi:2011qd}, which were obtained via ray-tracing through universes generated with high-resolution $N$-body simulations and gives $c = 0.16$ $\beta =  0.73$, $\alpha =  1.91$. 

Fig.~\ref{fig:sigmaD} also illustrates the impact of ellipticities, bias, and filaments on the weak lensing scatter, emphasizing that including filaments is particularly important. We note, however, that our current modeling of both filaments and halos, as well as the linear bias, remains relatively simplistic. A more refined treatment could plausibly bring our results into closer agreement with $N$-body simulations. We leave these refinements for future work. The primary goal of this study is to demonstrate the potential of weak lensing of bright  sirens and to motivate further improvements.

\section{Hubble diagram reconstruction}

We generate benchmark catalogues similar to those considered in previous studies (e.g.~\citet{Mangiagli:2023ize,Mpetha:2024xiu}). Specifically, we construct one catalogue consisting of $300$ binary neutron star (BNS) systems and another consisting of $12$ binary massive black hole (BMBH) systems. The luminosity distances of the BNS population are assumed to be measurable from the GW signals by ET, while those of the BMBH population are assumed to be measurable by LISA. The redshifts are assumed to be obtained from EM counterparts. BNS mergers are expected to produce multiple types of detectable EM counterparts, as demonstrated by GW170817~\citep{LIGOScientific:2017adf}. Finding $\mathcal{O}(100)$ bright BNS events over a 10 year period is considered realistic~\citep{Iacovelli:2022bbs,Loffredo:2024gmx}. In contrast, EM counterparts of BMBHs remain uncertain. However, several studies suggest that a small number of such events may be detectable~\citep{Dotti:2006zn,Tamanini:2016zlh,Mangiagli:2022niy,Mangiagli:2023ize}.

\begin{figure}
\centering
\includegraphics[width=\columnwidth]{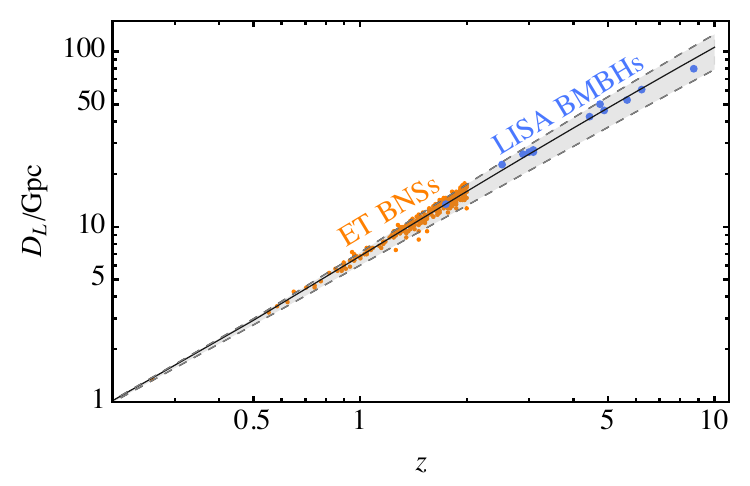}
\vspace{-4mm}
\caption{The solid curve shows the true luminosity distance-redshift relation $\tilde D_L(z)$, while the surrounding gray band indicates 99\% range of the apparent luminosity distances. The points represent a benchmark catalogue of 500 NS binaries and 10 MBH binaries.}
\label{fig:catalogue}
\end{figure}

The redshifts of the BNSs are drawn from the probability distribution
\be
    p_z(z) \propto (1+z)^{2.7}\,\frac{\td V_c}{\td z} \,,
\ee
where $V_c$ denotes the comoving volume. The power-law index $2.7$ is motivated by studies of the star formation rate density~\citep{Madau:2014bja} and is consistent with the LVK results~\citep{LIGOScientific:2025pvj}. The redshifts of the BMBHs are drawn from a population generated using a merger rate derived from the extended Press-Schechter formalism combined with the MBH-galaxy co-evolution model of~\citet{Urrutia:2024hwc}. We restrict the BNS population to $z_s < 2$ and the BMBH population to $z_s < 10$ (see e.g.~\citet{Mangiagli:2023ize}). For both populations, we assume negligible uncertainty in the redshift measurements. For the apparent luminosity distances, we adopt a relative measurement uncertainty of $\sigma_{D_L} = 0.03 D_L$ for the BNSs and $\sigma_{D_L} = 0.003 D_L$ for the BMBHs.\nobreakfootnote{These measurement uncertainties are intentionally optimistic compared to population averaged forecasts, presented for BNSs, for example, by~\citet{Iacovelli:2022bbs,Loffredo:2024gmx}, and for BMBHs by~\citet{Mangiagli:2023ize}, as they correspond to events for which an EM counterpart is identified. In such cases, information on the sky location and binary inclination from the EM measurement can substantially reduce the distance-inclination degeneracy in the GW signal~\citep{Holz:2005df,Shah:2013ema,Hotokezaka:2018dfi,Chen:2018omi,Nakar:2020pyd,deSouza:2023gjv}.} The catalogues are generated by sampling source redshifts from the respective redshift distributions and drawing a lensing magnification factor $\mu$ from the corresponding magnification PDFs. The resulting benchmark catalogues are shown in Fig.~\ref{fig:catalogue}.

To reconstruct the Hubble diagram, we perform a Markov chain Monte Carlo (MCMC) inference of the model parameters $\vect\theta = \{ \Omega_M, h, \sigma_8 \}$ by sampling the likelihood
\be \label{eq:likelihood}
    \mathcal{L}(\vect\theta) = \prod_{j=1}^{N} \frac{\int \td D_L \, p_{\rm obs}(D_{L,j}|D_L)  p_{\rm mod}(D_L|z_j,\vect\theta)}{P_{\rm det}(z_j,\vect\theta)} \,,
\ee
where $p_{\rm obs}$ is a Gaussian distribution with variance $\sigma_{\rm D_L}$ and
\be \label{eq:Pdet}
	P_{\rm det}(z,\vect\theta) \!= \!\! \int \!\td D'_L  \!\int  \td D_L \, p_{\rm obs}(D'_L|D_L) p_{\rm mod}(D_L|z,\vect\theta)
\ee
is the probability of observing a source at redshift $z$ in model $\vect\theta$. We assume that uncertainties in redshift measurements are negligible and that the detectability threshold corresponds to a cutoff in redshift rather than in luminosity distance. Under these assumptions, observational selection effects do not bias the inference. The probability distribution of the apparent luminosity distances at redshift $z$ is obtained from the true luminosity distances $\tilde D_L(z)$ and the weak lensing magnification distribution $p_\mu(\mu|z)$ as
\bea
	&p_{\rm mod}(D_L|z,\vect\theta) \\
    &\quad = p_z(z)  \int \! \td \mu \, p_\mu(\mu|z,\vect\theta) \delta\!\left[ D_L - \frac{\tilde{D}_L(z,\vect\theta)}{\sqrt{\mu}} \right] .
\eea
We use the delta function to perform the integrals over $D_L$ in \eqref{eq:likelihood} and \eqref{eq:Pdet}. If we approximated the weak lensing scatter in the apparent luminosity distances by a Gaussian distribution, the integral over $D_L$ in \eqref{eq:likelihood} and \eqref{eq:Pdet} would result in a Gaussian distribution with variance $\sigma_{\rm WL}^2 + \sigma_{D_L}^2$. In this work, we instead employ the full non-Gaussian magnification distribution.

A single likelihood evaluation, including the computation of the halo and filament mass functions and the weak-lensing distributions at multiple source redshifts, takes approximately 20~seconds on an Apple M1 processor. We sample the likelihood~\eqref{eq:likelihood} using the Metropolis-Hastings algorithm with Gaussian proposal distributions, whose widths are chosen to achieve an acceptance rate of $10-28\%$. We generate four Markov chains, each containing 2600 samples, of which the first 200 samples are discarded as burn-in. We adopt uniform priors in the ranges $h \in [0.59, 0.76]$, $\Omega_M \in [0.15, 0.47]$, and $\sigma_8 \in [0.4, 1.4]$. To assess the convergence, we employ the Gelman-Rubin statistic $\hat{R}$ and find acceptable convergence with $\hat{R} < 1.05$ for all parameters.

\begin{figure}
\centering
\includegraphics[width=\columnwidth]{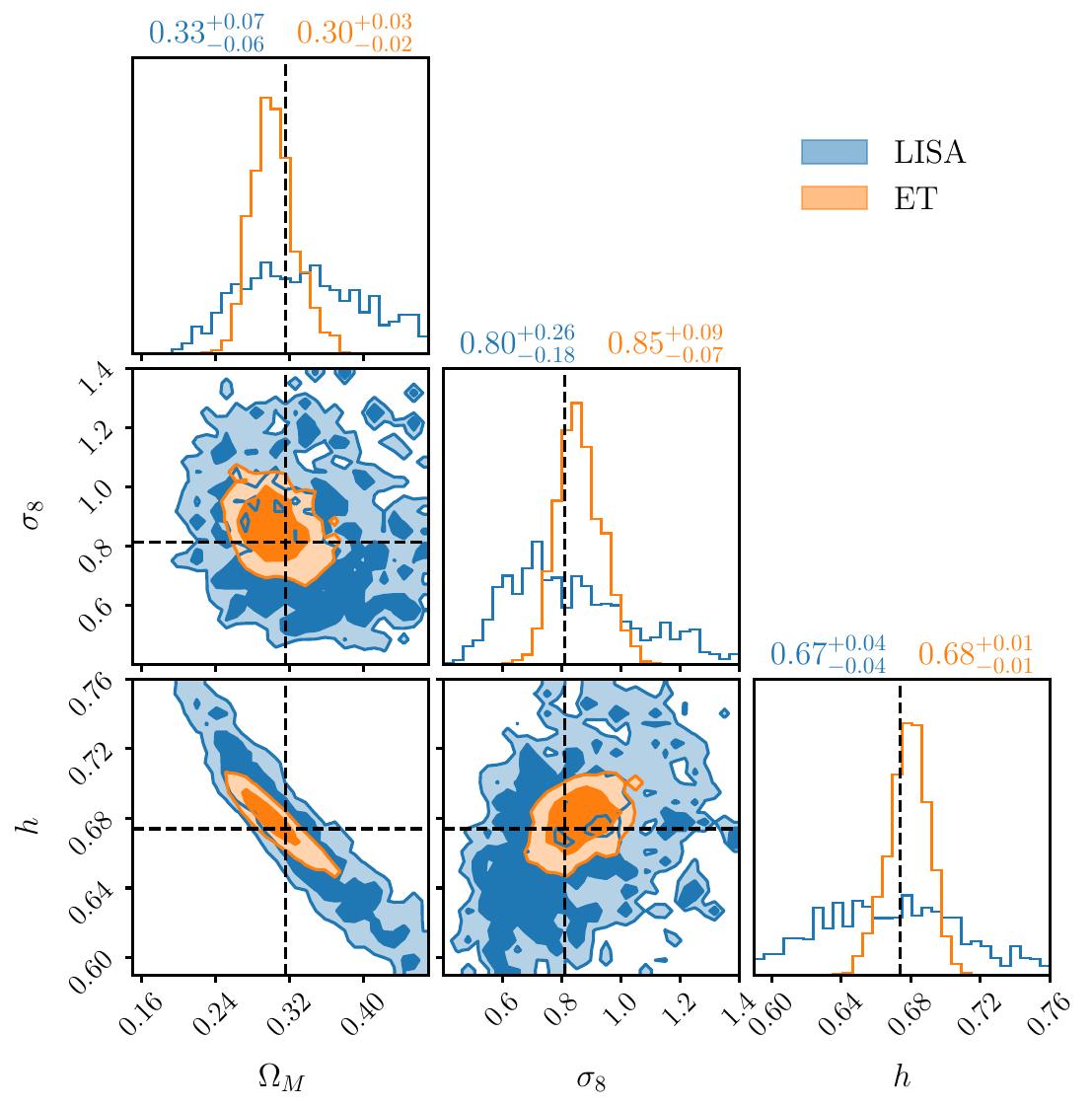}
\caption{Posteriors of the Hubble diagram reconstruction from the catalogues shown in Fig.~\ref{fig:catalogue}. The contours indicate the 68\% and 95\% credible regions. The dashed lines indicate the benchmark parameter values used to generate the catalogues.}
\label{fig:corner}
\end{figure}

In Fig.~\ref{fig:corner} we show the posterior distributions obtained for ET and LISA using the benchmark catalogues from Fig.~\ref{fig:catalogue}. The posteriors are consistent with the fiducial values (dashed lines), demonstrating that the inference is unbiased. ET provides significantly tighter constraints than LISA, despite LISA observing more distant sources. This difference is driven by the larger expected catalogue size for ET. In the $\Omega_M - h$ panel, a pronounced negative correlation is evident, indicating that lower Hubble constant values can be partially compensated by higher matter density. In contrast, $\sigma_8$ shows much weaker correlations with the other parameters. Quantitatively, we find that ET can measure $\sigma_8$ with an accuracy of 10\% and LISA with an accuracy of $30\%$, while their combination would yield an accuracy of $8\%$. The prospects for $h$ and $\Omega_M$ are similar to previous studies (e.g.~\citet{Mpetha:2024xiu}).

\section{Conclusions}

As GW detector sensitivities and EM follow-up capabilities continue to improve, bright standard sirens will become an increasingly important probe of cosmology. We have shown that weak-lensing-induced scatter in the luminosity distance-redshift relation encodes valuable information about the distribution of cosmic structures, enabling constraints on parameters that do not directly affect the mean Hubble diagram. Focusing on the amplitude of matter fluctuations, $\sigma_8$, we find that ET could achieve $10\%$ measurement with a sample of 300 neutron star binaries with identified EM counterparts. At lower GW frequencies, observations by LISA of massive black hole binaries provide a complementary avenue, yielding constraints at the $30\%$ level with a smaller sample of only 12 events.

Our results highlight that weak lensing transforms bright sirens from purely geometric probes into tracers of cosmic structures, opening a new channel for studying cosmological perturbations with GWs. While the precision of constraints from CMB observations~\citep{Planck:2018vyg} or from galaxy clustering and weak-lensing surveys~\citep{DES:2025tna} is unlikely to be reached, weak lensing of bright sirens provides a complementary probe of $\sigma_8$. Moreover, standard candles and standard sirens are uniquely sensitive to small-scale matter fluctuations that are largely inaccessible to conventional CMB or galaxy weak-lensing observations. In this work, we have focused on $\sigma_8$, but we plan to use bright-siren lensing to directly constrain the small-scale matter power spectrum.

Several avenues for future work naturally follow from this study. On the modeling side, improving the treatment of weak lensing through more realistic descriptions of halos, filaments, and clustering will be essential for reliable inference of cosmological parameters from bright siren data. From a methodological perspective, machine-learning-based approaches could significantly accelerate the inference pipeline and enable rapid marginalization over complex lensing models. This, in turn, would allow for a systematic exploration of how parameter uncertainties scale with the size and redshift distribution of the siren catalog, which will be crucial for assessing the constraining power of bright sirens.

The analysis can be readily extended to additional cosmological parameters beyond those considered in this study. In particular, including parameters such as the spectral tilt and spatial curvature would lead to more robust and realistic forecasts~\citep{Congedo:2018wfn}. Furthermore, exploring scenarios beyond the standard cold dark matter paradigm would be of interest. Since alternative dark matter models generally predict different abundances and internal properties of cosmic structures, the resulting weak-lensing-induced scatter of bright sirens could provide novel and complementary constraints on dark matter properties. For example, warm or fuzzy dark matter suppresses small-scale structures~\citep{Hu:2000ke,Bode:2000gq}, thereby reducing the weak-lensing-induced scatter, whereas compact dark matter objects would enhance the high-magnification tail~\citep{Zumalacarregui:2017qqd}.

\section*{Acknowledgments} 
We are grateful to G.~Congedo, K.~Kainulainen and J.~Urrutia for valuable discussions. This work was supported by the European Union's Horizon Europe program under the Marie Sk\l{}odowska-Curie grant agreement No. 101065736, the Estonian Research Council grants TARISTU24-TK3 and TARISTU24-TK10, and the Center of Excellence program TK202 of the Estonian Ministry of Education and Research.

\section*{Data availability}
The data generated for this article are available upon request from the author. The code used to generate the data is publicly available at~\href{https://github.com/vianvask/halos}{https://github.com/vianvask/halos}.

\bibliographystyle{mnras}
\bibliography{refs}

\label{lastpage}

\end{document}